\documentstyle[11pt,aaspp4]{article}
\begin{document}
\title{DETECTION AND CHARACTERIZATION OF COLD\\  INTERSTELLAR DUST AND PAH EMISSION,\\ 
 FROM  $\it COBE$ OBSERVATIONS}
\author{E. Dwek\altaffilmark{1}, R. G. Arendt\altaffilmark{2},  D. J. 
Fixsen\altaffilmark{2}, T. J. Sodroski\altaffilmark{3}, N. Odegard\altaffilmark{2}, J. L.
Weiland\altaffilmark{2}, W. T. Reach\altaffilmark{4}, M. G. Hauser\altaffilmark{5}, T.
Kelsall\altaffilmark{1}, S. H. Moseley\altaffilmark{1}, R. F. Silverberg\altaffilmark{1},
R. A. Shafer\altaffilmark{1}, J. Ballester\altaffilmark{6}, D. Bazell\altaffilmark{7}, and
R. Isaacman\altaffilmark{7}}

\altaffiltext{1}{Laboratory for Astronomy and Solar Physics, Code 685,  NASA/Goddard
Space Flight Center, Greenbelt, MD 20771.\ \ eli.dwek@gsfc.nasa.gov}
\altaffiltext{2}{ Hughes STX, Code 685.9, NASA/GSFC, Greenbelt, MD 20771}
\altaffiltext{3}{ Applied Research Corporation, Code 685, NASA/GSFC, Greenbelt, MD
20771}
\altaffiltext{4}{ Institut d'Astrophysique Spatiale, Batiment 121, Universite
Paris XI, 91405 Orsay cedex, France}
\altaffiltext{5}{Space Telecsope Science Institute, 3700 San Martin Drive, Baltimore, MD
21218}
\altaffiltext{6}{Emporia State University, 1200 Commercial, Emporia, KS 66801}
\altaffiltext{7}{ General Science Corporation, Code 902.3, NASA/GSFC, Greenbelt, MD
20771}

\begin{abstract} 
Using data obtained by the DIRBE instrument on the 
{\it COBE}\ \footnote{The National Aeronautics and Space Administration at Goddard
Space Flight Center (NASA/GSFC) is responsible for the design, development, and operation
of the {\it Cosmic Background Explorer (COBE)}. Scientific guidance is provided by the
COBE Science Working Group. GSFC is also responsible for the development of the analysis
software and for the production of mission data sets.}
spacecraft, we present the mean $3.5-240\ \mu$m
spectrum of high latitude dust. Combined with a spectrum obtained by the FIRAS
instrument, these data represent the most comprehensive wavelength coverage of dust in 
the diffuse interstellar medium, spanning the $3.5-1000\ \mu$m wavelength regime. At
wavelengths shorter than $\sim$ 60 $\mu$m the spectrum shows an excess of emission over
that expected from dust heated by the
local interstellar radiation field and  radiating at an equilibrium temperature. The
DIRBE data thus extend the observations of this excess, first detected by the IRAS
satellite at 25 and 12
$\mu$m, to shorter wavelengths. The excess emission arises from very small dust particles
undergoing temperature fluctuations. However, the  3.5-to-4.9 $\mu$m intensity ratio {\it
cannot} be reproduced by very small silicate or graphite grains. The DIRBE data
strongly suggest that the 3.5 - 12 $\mu$m emission is produced by carriers of the
ubiquitous 3.3, 6.2, 7.7, 8.6, and 11.3
$\mu$m solid state emission features that have been detected in a wide variety of
astrophysical objects. The cariers of these features have been widely identified with
polycyclic aromatic hydrocarbons (PAHs). 

Our dust model consists of a mixture of
PAH molecules and bare astronomical silicate and graphite grains with optical properties
given by Draine \& Lee. We obtain a very good fit to the DIRBE spectrum, deriving the size
distribution, abundances relative to the total hydrogen column density, and relative
contribution of each dust component to the observed IR emission. At wavelengths above 140
$\mu$m the model is dominated by emission from
$T\approx 17-20\ K$ graphite, and $15-18\ K$ silicate grains. The model provides a good fit
to the FIRAS spectrum in the $140-500\ \mu$m wavelength regime, but leaves an excess
Galactic emission component at $500-1000\ \mu$m. The nature of this component is still
unresolved.

We find that (C/H) is equal to $(7.3\pm 2.2)\ 10^{-5}$ for PAHs, and equal to 
$(2.5\pm 0.8)\ 10^{-4}$ for graphite grains, requiring about 20\% of the cosmic abundance of
carbon to be locked up in PAHs, and about 70\% in graphite grains (we adopt
$(C/H)_{\odot}=3.6\ 10^{-4}$). The model also requires all of  the available magnesium,
silicon, and iron to be locked up in silicates. The power emitted by PAHs is $1.6\ 10^{-31}\ 
W/H\ atom$, by graphite grains
$3.0\ 10^{-31}\ W/H\ atom$, and by silicates $1.4\ 10^{-31}\ W/H\ atom$, adding up to a
total infrared intensity of  $6.0\ 10^{-31}\ W/H\ atom$, or $\sim 2\
L_{\odot}/M_{\odot}$. 

The [C II] $158\ \mu$m line emission detected by the
FIRAS provides important information on the gas phase abundance of carbon in the diffuse
ISM. The 158 $\mu$m line arises predominantly from the cold neutral medium (CNM), and
shows that for typical CNM densities and temperatures
$C^+/H=(0.5-1.0)\ 10^{-4}$, which is $\sim 14-28$\% of the cosmic carbon abundance. The
remaining carbon abundance in the CNM, which must be locked up in dust, is about equal to
that required to provide the observed IR emission, consistent with notion that most ($\gtrsim
75$\%) of this emission arises from the neutral component of the diffuse ISM. 

The model provides a good fit to the general interstellar extinction curve. However, at
UV wavelengths it predicts a larger extinction. The excess extinction may be the result
of the UV properties adopted for the PAHs. If real, the excess UV extinction may
be accounted for by changes in the relative abundances of PAHs and carriers of the  2200
\AA\ extinction bump. 

\end{abstract}


\section{INTRODUCTION}

A surprising result of the {\it Infrared Astronomical Satellite
(IRAS)} all-sky survey was  the detection of a "new" high-latitude diffuse infrared
emission component dubbed "cirrus" because its morphology resembled that of filamentary
clouds in our own atmosphere (\cite{low84}). This IR emission component was not entirely
new, and had been previously detected  through scattered light at optical wavelengths since
the 1970's (e.g. \cite{dvf72}, \cite{san76}). The surprising aspect of the infrared (IR)
observations was their spectra which displayed an excess of emission at 12 and 25 $\mu$m
over that expected from dust particles radiating at the equilibrium temperature and
heated by the local interstellar radiation field (LISRF). Draine
\& Anderson (1985) attributed this excess to the effect of temperature fluctuations
in small grains.  The grain size distribution given by Mathis, Rumple, \& Nordsieck
(1977; hereafter MRN) was therefore required to  be extended to very small grain sizes of
$\sim 3$ \AA\  (\cite{da85}, \cite{weil86}). The cirrus spectrum may, however, arise from
the same population of dust particles that give rise to the "unidentified" IR emission
features at 3.3, 6.2, 7.7, 8.6, and 11.3 $\mu$m. These features were detected in a wide
variety of astrophysical objects, reflection nebulae, H II regions, and planetary nebulae
(see \cite{coh86} and references therein), and the large-scale emission in the 3.3 $\mu$m
band was detected by Giard et al. (1988, 1994). However, so far there has been no direct
detection of these features in high latitude cirrus clouds, a situation that may change when
analysis of data from the ISO and IRTS satellite missions are completed.

The carriers of these features are most commonly identified with a  class of molecules
called polycyclic aromatic hydrocarbons, or PAHs (\cite{dw81}, \cite{lp84},
\cite{atb85}, \cite{plb85}). However, that identification is not unique, and various
other compounds such as hydrogenated amorphous carbon (HACs; e.g. Duley 1989, and
references therein), or quenched carbonaceous composites (QCCs; e.g. Sakata \& Wada 1989,
and references therein) have been suggested as alternative carriers of these features.
Specifically, Papoular et al. (1989) suggested demineralized coal or vitrinite for the
carriers. The optical constants of $10-100$ \AA\ coal grains are such that the grains can
maintain {\it equilibrium} temperatures with near-IR colors in the $500 - 1000\ K$ range,
provided that they are exposed to radiative fluxes of the order of
$10^{-2}-1\ W\ m^{-2}$ (\cite{gui94}). These radiative fluxes are a few orders of
magnitude larger than those of the LISRF, suggesting that coals are not likely carriers
of the emission features in the diffuse interstellar medium (ISM). Throughout  our
analysis we will assume that PAHs are the carriers of the unidentified IR emission bands. 

Observations of cirrus made by the Diffuse Infrared Background Experiment (DIRBE) and Far
Infrared Absolute Spectrophotometer (FIRAS) experiments on board the {\it
Cosmic Background\ Explorer (COBE)} satellite (\cite{bog92}) represent the first spectra of
these objects  spanning the entire 3.5 to 1000 $\mu$m wavelength region. They therefore
provide the strongest constraints on the  nature of the dust in these regions. Bernard et
al. (1994, 1996) presented the DIRBE observed cirrus spectrum for low Galactic latitude
($|\it b|\le 10^o$). In this paper we complement their studies by deriving an average
high-latitude cirrus spectrum and extending the wavelength coverage to 1000 $\mu$m. We
also present the spectrum of select individual high-latitude cirrus clouds in order to
examine the spectral variations in these objects.  

In general, the determination of the cirrus spectrum from the DIRBE maps is complicated
as a result of the presence of strong foreground emissions from the interplanetary dust
cloud and at 3.5 and 4.9
$\mu$m from Galactic stellar emission. In
\S 2 we describe the data set used in the analysis, and briefly reiterate the method
developed by Arendt et al. (1996; also presented by \cite{was96}) for the extraction of
the cirrus spectrum. The IRAS data provided clear  evidence for the presence of
transiently heated dust particles in cirrus clouds (\cite{da85}). However, as a result
of the limited wavelength coverage of the IRAS ($\lambda = 12,\ 25,\ 60,$ and $100\
\mu$m), the cirrus data could not constrain the $\it {composition}$ of the very small
dust particles. In \S 3 we show that any extension of the MRN size distribution to very
small grains is inconsistent with the DIRBE data. Adopting the PAH model described by
D\'esert, Boulanger, \& Puget (1990) to characterize the carriers of the $3.3-11.3\ \mu$m
features (see Appendix A.2), we show that the DIRBE 3.5 to 25 $\mu$m data are consistent with
the predicted spectrum  from these macro-molecules. A detailed description of the dust model
used in this paper is presented in \S 4, and in Appendix A.1 we describe the method used to
calculate the PAHs heat capacities used in this paper. In
\S 5 we describe the fitting procedure, and discuss the uniqueness of the various model
parameters derived from fitting the DIRBE observations. We also calculate the abundances of
the various dust constituents relative to hydrogen, and calculate the extinction predicted by
the model, comparing it to the observed average interstellar extinction curve. The [C II] $158\
\mu$m line emission detected by the FIRAS instrument is used to calculate the gas
phase abundance of carbon in the diffuse ISM (Appendix A.3), which provides an independent
estimate of the amount of carbon required to be locked up in dust. In
\S 6 we compare the far-IR spectrum predicted by the model to the FIRAS observations. The
results of our paper are briefly summarized in \S 7.


\section{THE DERIVATION OF THE CIRRUS SPECTRUM} A description of the derivation of the
dust spectrum in the general ISM was presented by Weiland, Arendt, \& Sodroski (1996),
based on the detailed work of Arendt et al. (1996). The data used in the derivation
consists of the {\it COBE}/DIRBE sky maps of emission at $\lambda$ = 1.25, 2.2, 3.5, 4.9,
12, 25, 60, 100, 140, and 240 $\mu$m (with corresponding bandwidths of 570, 225, 220,
82.1, 135, 41.0, 23.2, 9.74, 6.17, and 4.96, in units of $10^{11}$ Hz), from which the
contributions of interplanetary dust particles and of discrete bright
stellar and extragalactic sources and unresolved Galactic starlight has been removed
(\cite{k96}, \cite{ar96}). The remaining intensity consists therefore of IR emission from
Galactic interstellar dust, and from a diffuse extragalactic background which is assumed
to be spatially isotropic. 

The spectrum of the ISM, relative to the $I(100\ \mu$m) intensity, was obtained by deriving
the slope of the $I(\lambda)$ versus
$I(100\ \mu$m) correlation for the high latitude regions of the sky that are unaffected
by residual interplanetary dust emission. The slope of the correlation is insensitive to
any isotropic terms, and the spectrum is therefore free of any extragalactic or
any isotropic Galactic emission components. The spectrum was then scaled to the Galactic
H I column density by correlating the $I(100\ \mu$m) intensity to N(H I) over the same
regions of the sky. However, by construction, the resulting IR emssion does not originate
only from the H I gas, but from all gas phases that correlate with the $I(100\ \mu$m)
intensity. 

At wavelengths $\lambda \le 4.9\ \mu$m the
subtraction of the stellar emission component is not sufficiently accurate to bring out
the dust emission component in the residual intensity maps. Therefore, Arendt et al. used
near-IR reddening-free colors to identify this component using DIRBE 1.25, 2.2, 3.5, and
4.9 $\mu$m data. The procedure is equivalent to identifying dust emission components in
color-color plots as the points that are offset from the stellar reddening curve. The
method could only be applied to Galactic latitudes with $\vert{\it b}\vert \le \ 25^o$
because of the presence of residual artifacts from the interplanetary dust emission at
higher latitudes. As a result there may be a systematic offset between the
the 3.5 and 4.9 $\mu$m intensities and the rest of the spectrum, if the ratio of the 3.5
or 4.9 to 100 $\mu$m intensity is a function of latitude.

The cirrus spectrum at wavelengths longward of 100 $\mu$m was derived from the FIRAS
Pass 3 data (see Fixsen 1994) from which the microwave background and the dipole have
been subtracted, in a fashion similar to that used with the DIRBE data at
$\lambda \ge\ 12\ \mu$m. The first step in this derivation consisted of degrading the
Galactic component of the DIRBE 100 $\mu$m emission to the FIRAS $7^o$ beam resolution
using the actual FIRAS beam pattern derived from observations of the moon. The full
FIRAS high frequency data set was used for 100 $< \lambda(\mu m)\ <$ 500, while the
combined low frequency data set was used for the $500-5000\ \mu$m wavelength region. The
spatial correlation with the DIRBE 100 $\mu$m intensity map was performed for each of the
167 FIRAS frequency channels. In both frequency regimes, the data were limited to $|b| >
45^o$ and weighted by the FIRAS weights. In the resulting spectrum, the FIRAS intensity
matches that of the DIRBE 240, 140, and 100
$\mu$m in the respective wavelength bands, confirming the consistency of the data sets and
the derivation procedure. Only the $140 -1000\ \mu$m spectrum of the FIRAS is presented
here since the data are too noisy in the remaining frequency channels.

The derived $3.5-1000\ \mu$m cirrus spectrum is therefore an
{\it average} spectrum of high-latitude interstellar dust. It represents the average
emission from dust integrated over all the non-isotropic ISM components. Dust residing
in H II as well as H$_2$ gas can therefore contribute to the emission. To examine the gas
phase dependence of the dust emission we derived the IR spectra of two distinct
 high latitude clouds. The first cloud (referred herafter as Cloud 1) corresponds to
cloud numbers MBM 53, 54, 55 in the catalog of high-latitude molecular cirrus clouds
(\cite{mbm85}). It represents a cloud in which the IR emission is dominated by dust
residing in the molecular gas and its IR emission is modeled in detail by Dwek, Arendt,
Fixsen, \& Reach (1996). The second (Cloud 2) was referred to as Cloud A by Low et al.
(1984) and  modeled by Draine \& Anderson (1985). This cloud is notable for having the
highest column density of any feature in the $b < -60^o$ H I map of Heiles (1975).  

Figure 1 depicts  the various cirrus spectra derived in this paper. The diamonds
represent the average ISM spectrum, and the spectra of Clouds 1 and 2 are
represented by triangles and stars, respectively. The average
ISM spectrum is normalized to a value of $0.7\ MJy\ sr^{-1}/N_H(10^{20}\ cm^2)$  at
$\lambda = 100\ \mu$m. The other spectra are offset by an arbitrary factor for sake of
clarity. Uncertainties in the data are determined by the RMS variation of the
$I(\lambda)/I(100\ \mu$m) colors in different regions of the sky, and equal to 20\%
(1 $\sigma$). We note that the average ISM spectrum presented in the figure is very similar
to that derived by Bernard et al. (1994, 1996).

Infrared lines
can contribute to the observed emission in the broad DIRBE bands. The most important
potential contributor in the diffuse ISM is line emission from [C II]
$158\ \mu$m in the
$140\ \mu$m DIRBE band. Our calculations (see \S 5.2 below) show that this line contributes
at  most
$\sim 2 \%$ to the emission in this band. We therefore attribute the observed spectra to
thermal emission from  interstellar dust. 

Table 1 summarizes the various spectra derived in this paper. Also presented in
the table are the dust color temperatures determined from the flux ratios in 
adjacent bands. Refering to the average cirrus spectrum, the 140-to-240
$\mu$m and the 60-to-100
$\mu$m flux ratios correspond to color temperatures of $\sim\ 19$ K, and $\sim\ 23$ K,
respectively (for a $\lambda^{-2}$ emissivity law), whereas the 12-to-25 $\mu$m and the
3.5-to-4.9 $\mu$m flux ratios give color temperatures of $\sim\ 210$ K, and $\sim\ 550$ K,
respectively.

The excess emission at shorter wavelengths has been attributed to the fact
that the cirrus emission arises from two distinct populations of dust particles
(\cite{da85}). The first consists of particles radiating at a relatively narrow range of
equilibrium temperatures between 16 and 20 K, and the second consists of dust particles
small enough to undergo temperature fluctuations. Their contribution to the emission is
first noticable at $60\ \mu$m, as manifested in the constancy of the 60-to-100
$\mu$m Galactic IRAS flux ratio with galactic longitudes   (\cite{s89}, \cite{s94} ),
 and increases at shorter wavelengths to dominate the emission at wavelengths below $25\
\mu$m. Comparison of the spectra depicted in Figure 1, or listed in Table 1, shows
variations in the long and short wavelength color temperatures. These variations probably
reflect differences in the particle size distributions and/or variations in the intensity
or spectrum of the interstellar radiation field.  An excellent review of the conditions
leading to the stochastic heating of interstellar dust particles and the observational
effects of the resulting temperature fluctuations is given by Aannestad (1989).   
\section{THE DIRBE EVIDENCE FOR THE PRESENCE OF PAHs} 
The IRAS observations of the diffuse ISM were of limited wavelength coverage and could
therefore not constrain the composition of the very small dust particles. Models by Draine
\& Anderson (1985) and  Weiland et al. (1986) could fit the IRAS fluxes by simply
extending the size distribution in the interstellar graphite-silicate MRN dust model to
very small grain sizes ($a\approx 3$ \AA\ ). Subsequent calculations (\cite{gd89})  have shown
that graphite and silicate grains with radii less than
$\sim$ 4 \AA, and $\sim 5$ \AA, respectively evaporate after residing for $\sim
4\times10^5$ yrs in the LISRF.  The extension to
grain sizes below these values is therefore unrealistic in the absense of any
replenishing mechanism. Similarly, only PAHs with more than
$\sim 20$ carbon atoms will survive evaporation in the LISRF (\cite{o86}).  

The DIRBE has more extensive wavelength coverage than the IRAS, and any extension of the
standard MRN bare silicate-graphite model to small grain sizes that survive evaporation
by the LISRF is inconsistent with the DIRBE data. However, the DIRBE 3.5 to 25
$\mu$m data are consistent with the predicted spectrum  from PAHs,  if we adopt the
 model described by D\'esert, Boulanger, \& Puget (1990) to characterize the carriers of
the $3.3-11.3\ \mu m$ features (see Appendix A.2). The "detection" of PAHs with the DIRBE
is demonstrated in Figures 2a$-$2d which compare various observed intensity ratios with
calculated in-band DIRBE intensity ratios of individual grain sizes for various grain
compositions. Figure 2a shows this comparison for the 3.5-to-4.9 $\mu$m intensity ratio
(hereafter designated as
$R_{3.5/4.9}$). The figure illustrates that no size distribution of silicate or graphite
grains can reproduce the observed intensity ratio, since even the smallest grains ($a$ = 4
\AA, for graphite, and $a$ = 5 \AA\ for silicates) have intensity ratios that fall below
the observed one. The  value of 
$R_{3.5/4.9}$ is expected to decrease with increasing grain size, since the amount of
short wavelength emission depends strongly the maximum temperature reached by the
stochastically-heated particles, and for a given radiation field, this temperature
decreases with increasing grain size. On the other hand, small PAHs {\it can} produce
larger values of $R_{3.5/4.9}$ because they posses a strong emission feature at 3.3 $\mu$m.
The figure also gives a good indication of the range of PAH sizes that is required to fit the
observed intensity ratio. 
PAHs are two-dimensional structures, and the relation between the PAH radius and its number
of carbon atoms, $N_c$, is given by $a$(\AA)=0.913$\sqrt{N_c}$ (D\'esert, Boulanger, \& Puget
1990).
The figure suggests an average PAH radius of $\sim$ 7 \AA, which consists of about $60$
carbon atoms. This average PAH size is in reasonable agreement with the value of $\sim 90$
carbon atoms estimated by  L\'eger, d'Hendecourt, \& D\'efourneau (1989) to comprise a typical
PAH molecule giving rise to the observed 3.3 to 11.3 $\mu$m color temperature.  Figure 2b
illustrates the same effect for $R_{12/25}$, the
$I(12\mu m)/I(25 \mu m)$ intensity ratio. In this case, very small graphite and silicate
grains can reproduce the observed intensity ratio, illustrating why a simple extension of
the standard MRN dust model to very small grain sizes could explain the IRAS data. Figures
2c and 2d continue this comparison for $R_{60/100}$ and
$R_{140/240}$, respectively, both covering wavelength regimes where dust radiating at the
equilibrium temperature dominates the emission. This fact is well illustrated in these
figures which show that PAHs have significantly higher color temperatures than those
implied by the observations. PAHs therefore can not be important contributors to the
emission at these wavelengths.
\section{A DESCRIPTION OF THE DUST MODEL} 
The IR flux from any given line of sight is the
sum of emissions from dust residing in  all the diffuse ISM phases along that line of
sight. The specific IR intensity  from a population of dust particles characterized by a
size distribution $f(a)$ in the radius interval
$a_{min}\le a
\le a_{max}$, and a dust-to-gas mass ratio $Z_d$, can therefore be written in the most
general form as:
\begin{equation} {I(\lambda)}={\mu
m_HZ_d\over{<m_d>}}\times{N_H}\int_{a_{min}}^{a_{max}}da\ f(a)\
\sigma(a,\lambda)\int dT\ {\cal P}(a,T)\ B_{\lambda}(T)	
\end{equation}
where $\mu$ is the mean atomic weight of the gas in {\it amu},
$N_H\equiv N_{H\ I}+N_{H\ II}+2N_{H_2}$ is the {\it total} H-column density along the
line of sight, $<m_d>\equiv\int_{a_{min}}^{a_{max}}da\ f(a) {4\pi\over3}\rho a^3$, is the
size-averaged mass of the dust population,  $\rho$ is the mass density of a dust particle,
$\sigma(a,\lambda)$ is its cross section at radius $a$ and wavelength
$\lambda$, $B_{\lambda}(T)$ is the Planck function, and ${\cal P}(a,T)dT$ is the
probability that the temperature of a dust particle will be between $T$ and $T+dT$. For
dust particles radiating at the equilibrium dust temperature,
$T_{eq}$, the function ${\cal P}(a,T)$ is simply a $\delta-$function at $T=T_{eq}$.
Implicit in equation (1) is a summation over the various grain compositions. A similar
equation applies for PAHs, except that we represent the PAH size distribution as a
function of the number of carbon atoms, rather than grain radius. 

The dust model used in this paper consists of a mixture of bare silicate and graphite
particles, and PAH molecules. An important physical property of the radiating dust
particles is their heat capacity. This quantity determines the temperature fluctuations,
and hence the function ${\cal P}(a,T)$ of the various dust particles. Appendix A.1 
describes the heat capacities used for the various grain constituents, with emphasis on
PAHs, for which we used the group additive method to determine their value. The relation
between
${\cal P}(a,T)$ and the various grain parameters and the LISRF has been described by
Aanestad (1989). The LISRF used in this paper was taken from Mathis, Mezger, \& Panagia
(1983). Also summarized in the Appendix are  the radiative cross sections used in the
model. Graphite and silicate optical properties were calculated from Mie theory using the
optical constants of Draine \& Lee (1984). Optical properties of the PAH were adopted
from the work of L\'eger, d'Hendecourt, \& D\'efourneau (1989), as summarized by
D\'esert, Boulanger, \& Puget (1990). 

Given the dust composition and physical and optical properties, a dust model must specify
the size distribution and relative abundance of its various constituents. The
distributions in grain radii, $a$, adopted for the various grain material are:
 
(a) for silicates:
\begin{equation} 
f(a) = a^{\gamma}			\qquad \qquad          a_{min} \le a \le a_{max}
\end{equation} 
with \{$a_{min},\ \gamma,\ a_{max}  $\} =
\{$0.0050\ \mu$m, $ -3.5$, $0.25\ \mu$m\}, their MRN value, except that the lower size
limit was extended from $0.025\ \mu$m,
 the nominal MRN value, to a lower value of $ 0.0050\ \mu$m;

(b) for graphite:
\begin{eqnarray}
f(a) &= a^{\gamma_1}\qquad \qquad & a_{min} \le a \le a_b \nonumber \\
&=a^{\gamma_2}\qquad \qquad & a_b \le a \le a_{max} 
\end{eqnarray}
allowing for a break in the slope of the power law at some intermediate
radius $a_b$; 

and (c) for PAHs:
\newline
The grain size distribution is written as a power law in $N_c$, the
number of carbon atoms in a PAH molecule:
\begin{equation} 
f(N_c) = N_c^{\gamma_p} \qquad \qquad          N_{c1} \le N_c \le N_{c2}
\end{equation} 
The number of carbon atoms in the smallest  PAH was fixed at $N_{c1}=20$
since smaller PAHs will evaporate when exposed to the LISRF (\cite{o86}). The number of
carbon atoms  characterizing the largest PAH, $N_{c2}$, and
the power law of the PAH size distribution, $\gamma_p$,  were taken to be free parameters
of the model.

We have considered two models using with different constraints on the
graphite size distribution. For both models the graphite size distribution is 
 characterized by \{$\gamma_2, a_{max}$\} = \{$-3.5$, $0.25\ \mu$m\} for radii above 
$a_b$. The models differ in their characterization of the distribution of very
small graphite grains. The first model (model A) considers the very small graphite grains
and the PAHs to be independent populations of particles. We adopt the MRN grain size
distribution for this model, but allow
$a_{min}$, the lower limit of the size distribution, to attain values below the nominal
MRN value of $0.0050\ \mu$m. In this model, $\gamma_1 = \gamma_2 = -3.5$, the
MRN value.  

In the second model (model B), the graphite size distribution is smoothly joined to that
of the PAHs, a distribution one would expect if PAHs were created from the destruction
and processing of graphite grains in the ISM.  The size of the smallest graphite grain
$a_{min}$ was therefore chosen so that the number of carbon atoms in a 
$\it {spherical}$ graphite particle of radius $a_{min}$ would equal $N_{c2}$, the number of
carbon atoms in the  largest PAH, that is,
\begin{equation} a_{min}(\mu m)=1.29\times10^{-4}\ N_{c2}^{1/3}
\end{equation} The power law of the size distribution in the \{$a_{min}, a_b$\} size
interval was chosen so that $dn(N_c)/dN_c$ is continuous  across the PAH-graphite
boundary at $a_{min}$. The value of $\gamma_1$ is therefore related to $\gamma_p$ by
\begin{equation}
\gamma_1=3\gamma_p+2
\end{equation} The constraints on $a_{min}$(graphite) and $N_{c2}$, and on $\gamma_1$ and
$\gamma_p$ as expressed in equations (5) and (6) are quite arbitrary, but they have the
aesthetic virtue of providing a smooth transition from graphite grains to PAHs.

The upper limit of the size distribution of the graphite and silicate grains is
uncertain. It cannot be constrained from extinction measurements, since large grains are
essentially gray particles. However, the far-IR emission is sensitive to the value of the
upper limit, since larger dust particles are colder. Unfortunately, it is difficult to
disentangle the effects of grain size from the effects of the intensity of the interstellar
radiation field. A higher value of the LISRF will increase the average dust temperature,
an effect that can also be achieved by reducing the value of $a_{max}$ by an appropriate
value. With this ambiguity in mind, we simply followed previous dust models
(e.g. \cite{da85}), and chose the value of this cutoff to be at $a_{max}=\ 0.25\
\mu$m.

To summarize, the dust model is characterized by 14 parameters: 4 for silicates
\{$a_{min},\ \gamma,\ a_{max},\ Z_{sil}$\}; 6 for graphite \{$a_{min}, \gamma_1,
a_b, \gamma_2,  a_{max},\ Z_{grf}$\}; and 4 for the PAHs \{$N_{c1},
\gamma_p,  N_{c2},\ Z_{PAH}$\}. Of these parameters, 8 are fixed, and 6 are
allowed to vary in the calculations.   Table 2 summarizes the various parameters of the 
dust model. 

\section{MODEL RESULTS}
\subsection{General} To find the dust model that best fits the observed
spectrum we mapped out the 3-dimensional $\chi^2$ space spanned by the model parameters:
\{$\gamma_p,\ N_{c2},\ a_{min}$(graphite)\}, or \{$\gamma_p,\ N_{c2},\ a_b$\} depending on the
characterization of the graphite grain size distribution (model A or B, respectively).
For each value of these 3 parameters, the PAH-to-gas mass ratio,
$Z_{PAH}$ was determined from fitting DIRBE 3.5, 4.9, and 12 $\mu$m observations with the
PAH spectrum. The graphite and silicate dust-to-gas mass ratios,
$Z_{grf}$ and $Z_{sil}$, respectively, were then determined from a $\chi^2$-fit of the
model to the residual spectrum. 

We expect some correlation between the various dust parameters. For example, $N_{c2}$ and
$\gamma_p$ should be strongly correlated as suggested by Figure 2a. The figure shows that
the observed $3.5$-to-$4.9\ \mu$m flux ratio is about equal to that produced by PAHs with
$\sim 60$ carbon atoms. The ratio can however be equally well reproduced by a range of
PAH sizes, from $N_{c1}=20$ to some upper limit $N_{c2}$. The larger the value of
$N_{c2}$ the less these PAHs will contribute to the 
$3.5\ \mu$m emission, and the more the spectrum will need to weighted towards the smaller
PAH sizes. Large values of $N_{c2}$ will therefore correspond to steep power laws, i.e.
large values of $\vert\gamma_p\vert$.

For a given spectrum, grain abundances are primarily determined by their 
opacities per unit mass. These quantities are relatively insensitive to
grain radius, so that the derived abundances are only sensitive to details of the IR
spectrum. For example, the abundance of silicates is
 sensitive to the $140$-to-$240\ \mu$m flux ratio. Exposed to the LISRF, 0.20
$\mu$m radius silicate grains attain a lower equilibrium temperature ($T_{eq} \approx 15\
K$) than identically sized graphite particles for which $T_{eq}\approx\ 18\
K$. Smaller values of $R_{140/240}$, will require colder dust to participate in
the emission, and will therefore lead to larger values of $Z_{sil}$.   

Table 3 lists the values of the parameters of four models for the average ISM that produced
the lowest values of $\chi^2$. For each choice of
the graphite grain size distribution (model A or B in Table 2),  the four models represent
the extreme values of $N_{c2}$ that still provide a good fit to the data. The various
models produce almost indistinguishable fluxes with minor differences in the spectra of
the various grain constituents. Typical
values of $N_{c2}$ are between $\sim$100 and 170. Figure 3 presents the spectrum
for dust model A with
$N_{c2}=100$. The figure shows that PAHs dominate the emission at  wavelengths $\lambda
\lesssim 20 \mu$m, whereas stochastically heated graphite grains are important
contributors to the emission in the narrow wavelength region of
$40 \lesssim \lambda(\mu$m$) \lesssim 70$. Graphite and silicate grains contribute about
equally to the emission at wavelengths above $\sim 300\ \mu$m. The main effect of the
silicate grains is to broaden the peak of the spectrum around $\sim 140\ \mu$m.
The power emitted by PAHs is $1.6\ 10^{-31}\  W/H\ atom$. Graphite grains emit
$3.0\ 10^{-31}\ W/H\ atom$, and silicates emit $1.4\ 10^{-31}\ W/H\ atom$. The
total infrared intensity is  $6.0\ 10^{-31}\ W/H\ atom$, or $1.9\
L_{\odot}/M_{\odot}$, compared to an average value of $\sim\ 3.0\ L_{\odot}/M_{\odot}$ 
found for the inner Galaxy (\cite{s94}). 

PAH size distributions derived here are very similar to that derived by D\'esert
et al. (1990). Translating the results from their Table 2 to parameters used in this paper
we get for their model: $N_{c1}=19;\ N_{c2}=170;\ \gamma_p=-1.5$. Similar results were
obtained by Siebenmorgen \& Kr\"ugel (1992) who derived a typical PAH
cluster size of 150 carbon atoms. 

\subsection{Dust and PAH Abundances}
Assuming that the H$_2$ and H$^+$ contributions to the total H-column density are
negligible, we get that the total dust-to-gass mass ratio of the model, $Z_{dust}$, is equal
to 0.0080, consistent with the value of 0.0073 allowed from cosmic abundance
considerations (\cite{ag89}). The small excess may be attributed to the fact that
$N_H$ may be larger than $N_{H\ I}$, which would indicate that dust in H I-correlated ionized
or molecular medium contributes to the IR emission. The size of the discrepancy seems, however,
to indicate that this contribution is small compared to that of the dust residing
in the neutral medium.  We also note $Z_{dust}$ may be larger than the quoted value, since an
additional mass of dust may be required to explain the cold dust component in the
FIRAS spectrum (see \S6 for more details), or may be present in the ISM but too cold to
manifest itself in the current observations.  

Similar results are obtained for the carbon abundance required by the model to be in
either the PAHs or the graphite grains. The mass fraction of carbon in PAHs is (see Table
3): $Z_{PAH}=0.00062$, (giving $C/H=7.3\ 10^{-5}$ for a mean molecular weight $\mu=1.42$),
and for graphite grains 
$Z_{grf}=0.0021,\ (C/H=2.5\ 10^{-4}$). The calculated carbon mass
fraction in the solid phase is thus $Z_{C(solid)}=0.0027\ (C/H=3.2\ 10^{-4})$, slightly
less than the cosmic abundance value of $Z_C(\odot)=0.0030\ [\ (C/H)_{\odot}=3.6\
10^{-4}]$ (\cite{ag89}).

The "missing" carbon must be in the gas phase and singly ionized (see discussion in
Appendix A.3). Its abundance can therefore be derived from the [C II] 158 $\mu$m line
emission detected by the FIRAS (\cite{w91}, \cite{ben94}). The $I(158\ \mu m)/I(100\ \mu
m)$ line ratio is latitude dependent,
 and subtracting the dust continuum emission we find a $C^+$ cooling rate per H-atom of
$1.45\ 10^{-33}\ W/H\ atom$ for the high-latitude ISM. This is about half the average value
of $2.65\ 10^{-33}\ W/H\ atom$ derived by Bennett et al. (1994) from data that
included lower latitude observations. The power in the 158 $\mu$m line constitutes about
2\% of the total energy received in the DIRBE 140 $\mu$m band, and its intensity implies a
[$C^+$]/[H I] ratio of
$\sim (0.5-1)\ 10^{-4}$  (see Appendix A.3), which is about equal to $14-28$\% of the cosmic
carbon abundance. The {\it total} carbon to H-atom ratio is therefore $C/H=(3.7-4.2)\ 10^{-4}$.
This value is consistent with the cosmic abundance limit, confirming previous results that
any contribution of the ionized gas to the IR emission is small ($\lesssim 15\%$).
Assuming that all the IR emission originates from the H I gas, we get that $\sim$ 18\% of
the carbon is locked up in PAHs, $\sim$ 62\% in graphite, and the remaining $\sim$20\% is
in the gas phase.

The abundance of PAHs needed to account for the IR emission can also be
estimated in the following two ways: the first by balancing the energy absorbed
by the PAHs from the LISRF to the {\it observed} emission in the $3.5-12\ \mu$m wavelength
regime, and independently, by calculating the PAH optical depth at some representative
wavelength dominated by PAH emission.

In the first method the fraction of carbon atoms locked up in PAHs is determined by
the ratio of the energy absorbed per C-atoms locked up in PAHs:
\begin{equation}
c{\int U_{LISRF}(\lambda)\ \tilde\sigma_{PAH}(\lambda) d\lambda}=2.6\ 10^{-27}\ W/ (C\
atom)
\end{equation}
and the observed IR luminosity emitted by the PAHs per H atom:
\begin{equation}
4\pi I_{PAH}/N_H=1.6\ 10^{-31}\ W/(H\ atom) 
\end{equation}
where in eq. (7), $U_{LISRF}$ is the energy density of the LISRF, and $\tilde\sigma_{PAH}$
is the PAH cross section per carbon atom in the molecule. The resulting value is:
\begin{equation}
(C/H)_{PAH} = 6.1\ 10^{-5}
\end{equation}
similar to the value of $(C/H)_{PAH} = 5.4\ 10^{-5}$ derived by Puget \& L\'eger
(1989). 

In the second method we calculate the abundance of PAHs required to account for the
observed IR emission. These calculations are a simple analytical representation of the
detailed model used to derive the results presented in Table 3. We first
assume that  all the intensity at a given wavelength
$\lambda$ originates from PAHs radiating at an {\it equilibrium } temperature
$T_{eq}$. The PAH optical depth $\tau_{PAH}$ per H column density will then be given by:
\begin{equation}
{\tau_{PAH}(\lambda)\over N_H}={I(\lambda)/N_H\over B_{\lambda}(T_{eq})}
\end{equation}
where $B_{\lambda}(T)$ is the Planck function. The  PAH optical depth can also be
expressed as:
\begin{equation}
\tau_{PAH}(\lambda)=N_C\ \tilde\sigma_{PAH}(\lambda)
\end{equation}
where $N_C$ is the column density of carbon atoms locked up in PAHs.
The C/H ratio locked up in PAHs is then derived from equations (7) and (8):
\begin{equation}
(C/H)_{PAH}={I(\lambda)/N_H\over \tilde\sigma_{PAH}(\lambda)\ B_{\lambda}(T_{eq})}
\end{equation}
We now assume that all the emission in the DIRBE $12\ \mu$m is emitted by PAHs. From
Table 1 we get that $I_{12\
\mu m}/N_H=3.16\ 10^{-16}\ Jy\ sr^{-1} cm^{-2}$. PAHs contributing to the emission will be
at temperatures of about 425 K, for which $B_{12\ \mu m}(425\ K)=1.46\ 10^{15}\ Jy\
sr^{-1}$, and the PAH cross section per unit C-atom is
$\tilde\sigma_{PAH}(12\ \mu m)=1.4\ 10^{-21}\ cm^{2}$. The resulting value of $(C/H)_{PAH}$ 
is $1.5\ 10^{-10}$.

A major assumption made in the derivation of this number is that the PAHs
radiate at an equilibrium temperature, when in fact, they are undergoing temperature
fluctuations. Therefore, at any given time, the $12\ \mu$m observations sample only a
fraction of the PAH population, those that happen to be at temperatures around 425 K. The
fraction of "flickering" PAHs is given by $\tau_{cool}(T)/<\tau_{abs}>$,
where
$\tau_{cool}(T)$ is the cooling time at temperature T, and $<\tau_{abs}>$ is an average
time between photon absorptions. The latter quantity can be simply calculated from:
\begin{equation}
<\tau_{abs}>^{-1}={\int ({\lambda \over h})U_{LISRF}(\lambda)\ \sigma_{PAH}(\lambda)
d\lambda}
\end{equation}
where $h$ is the Planck constant. The PAH cooling time is given by:
\begin{equation}
\tau_{cool}(T)^{-1} = {1\over T} {4<\tilde\sigma_{PAH}(T)>\sigma T^4\over C_{PAH}(T)}
\end{equation}
where $<\tilde\sigma_{PAH}(T)>=1.4\ 10^{-21}\ cm^2$ is the Planck-averaged value of the PAH
cross section per C atom, and $\tilde C_{PAH}(T)=2\ 10^{-16} erg\ K^{-1}$ is the PAH heat
capacity per C atom given by eq. (A-4). The resulting cooling time around T = 425 K is
$\sim 8\ s$. The absorption time is size dependent. To yield a $(C/H)_{PAH}$ value
consistent with the value obtained in eq. (9) the fraction of flickering PAHs must
be equal to $1.5\ 10^{-10}/6.1\ 10^{-5}=2.4\ 10^{-6}$, that is, the average time
between photon absorption should be $3.2\ 10^6\ s$. The required PAH size is
$N_c\approx 80$, consistent with the typical PAH sizes giving rise to the IR emission.

The derived fraction of carbon locked up in PAHs is about 2 times the value
derived by D\'esert et al. (1990; $C/H=3\ 10^{-5}$), Verstraete \& L\'eger (1992; $C/H=4\
10^{-5}$), or Siebenmorgen \& Kr\"ugel (1992;
$C/H=3\ 10^{-5}$), but comparable to that estimated by Puget \& L\'eger (1989; $C/H=5.4\
10^{-5}$). 
The abundance of the PAHs is directly proportional to $I_{PAH}(12\
\mu m)$/$N_H$, and inversely proportional to the wavelength integrated value of
$U_{LISRF}\times\tilde\sigma_{PAH}$. Differences between the PAH abundance derived
here and those derived by D\'esert et al. can be completely accounted for by the fact that the
value of the $\nu I_{PAH}(12\ \mu m)/\nu I(60\ \mu m)$ ratio in our spectrum is larger by a
factor of $\sim$ 1.45 than their value. Differences between our PAH abundance determination and
that of Verstraete \& L\'eger are due to their use of larger UV-visual cross-sections
for the PAHs.

Uncertainties in the derived PAH abundance are dominated by uncertainties in the value of the
LISRF, the $I_{PAH}(12\ \mu m)$/$N_H$ ratio, and the
UV-optical cross sections of the PAHs. Total uncertainties in the LISRF are about 15\%
(Mathis, Mezger, \& Panagia 1983). The uncertainty in the ratio of the 12 $\mu$m intensity to
H-column density ratio is determined by the combined uncertainties in the $I(100\ \mu
m)$/$N_H$ and $I(12\ \mu m)/I(100\ \mu m)$ ratios from which it was derived (Arendt et
al. 1996). Uncertainties in each of these quantities are about 10\% and 20\%, respectively.
Combined with the uncertainties in the LISRF, we estimate the uncertainty in the PAH abundance
to be about 30\%. Uncertainties resulting from those
in the UV-optical cross section of the PAHs are harder to evaluate. In our model we have used
 a value of $\sim 3\ 10^{-18}\ cm^2$/C\ atom in the 1000 - 3000 \AA\
wavelength range, which is a conservative value according to Allamandola, Tielens, \& Barker
(1989). Significantly larger UV absorption cross sections will reduce the required
PAH abundance. Laboratory data suggest that PAH absorption cross sections in this
wavelength region can be larger by about a factor of ten (Joblin, L\'eger, \& Martin 1992).
Such large cross sections will make them significant contributers to the UV extinction in
this wavelength regime, and potential carriers of the 2200
\AA\ extinction feature. This possibility will require significant modifications to current
dust models, and will not be further considered here. In summary, for the UV-optical cross
sections adopted in this paper, the estimated uncertainty in the PAH abundance is about 30\%.
The same uncertainty applies to the silicate and graphite abundances derived in this paper.

There has recently been an accumulating body of evidence suggesting that the carbon
abundance in the gas phase of the local ISM is $(C/H)_{ISM}= (2.25\pm 0.5)\ 10^{-4}$ (Snow \&
Witt 1995), about 60\% of its solar value. This lower value severely constrains the amount of
carbon that can be locked up in dust, leading various authors (Kim \& Martin 1996, Mathis
1996) to construct dust models that use up minimal amounts of carbon. The lower C/H
ratio in the local ISM is marginally consistent with the lowest value of $\sim 2.7\ 10^{-4}$
present in PAHs, graphite, and in the form of C$^+$ that is implied by our observations.

\subsection{Comparison to the General Interstellar Extinction}
An additional constraint
on the dust model is provided by the interstellar extinction. It is reasonable to expect
that the  extinction provided by the dust that fits the average high-latitude ISM emission
spectrum will also provide a reasonable fit to the average interstellar extinction curve.
Figure 4a compares the extinction from model A with $N_{c2}=100$ to the average
interstellar extinction curve of Mathis (1990). For sake of the comparison, we have
multiplied the observed extinction curve, which is normalized to the {\it total} H
column density, by the $N_H/N_{H I}$ ratio, which is equal to $\sim$1.2 (Savage \&
Mathis 1979). The figure shows that the predicted
extinction is similar to the average interstellar extinction for the H I gas.
 At wavelengths $\gtrsim\ 20\
\mu$m the value of the extinction is uncertain by at least a factor of two (Mathis 1990),
so that the model extinction is essentially consistent with the observations.

At UV wavelengths differences between the predicted and the average observed
interstellar extinction curve may reflect fundamental uncertainties in
dust models, or may suggest a real excess in the high latitude extinction compared to the
average Galactic value which is biased towards lower latitudes. The extinction predicted by
the model is higher by a factor of $\sim$ 1.3 at $\lambda=0.125\ \mu$m, increasing to a factor
of 2 at 0.1 $\mu$m (Figure 4b). The contribution of PAHs to the interstellar extinction is
quite uncertain (\cite{vl92}, \cite{jlm92}). The UV properties of
the PAHs derived by D\'esert et al. were tailored to fit the observed interstellar extinction
curve given in Savage \& Mathis (1979). As originally suggested by Greenberg \& Chlewicki
(1983), the carriers of the 2200
\AA\ extinction feature in the D\'esert et al. model have an otherwise flat extinction curve,
and represent a distinctly different population of particles from those responsible for the
rise in the FUV extinction. A similar approach was taken by Siebenmorgen
\& Kr\"ugel (1992), who adopted a small graphite grain component in their dust model, but
without the characteristic rise in extinction at UV wavelengths (Fig 1b in their paper),
attributing it entirely to PAHs.

The excess of high latitude UV extinction predicted in this paper is therefore directly
the result of the enhanced abundance of PAHs derived in our model (a factor of two
larger than that derived by D\'esert et al.) and the adopted PAH UV properties. In our
model, the 2200 \AA\ feature is provided by small graphite particles, which also contribute to
the rising extinction in the FUV (see also Draine \& Lee 1984, Figure 7). Clearly the UV
properties of the PAHs should be modified if graphite grains are allowed to contribute to
the UV extinction as well. A reduction in the rise of the UV absorptivity of the PAHs will
reduce the excess UV extinction exhibited in Figure 4b.

There is some observational evidence suggesting that the UV properties of the PAHs may
need to be modified. Boulanger, Pr\'evot, \& Gry (1994) have searched for a correlation
between the strength of the 12 and 25 $\mu$m IRAS emission and UV extinction from select lines
of sight through the Chamaeleon molecular cloud complex. They find a correlation between the
intensity of the 12 $\mu$m emission, which is a measure of the PAH abundance, and the strength
of the 2200 \AA\ extinction feature, suggesting a common carrier. Furthermore, they found no
correlation between the intensity of the 12 $\mu$m emission and the rise in the extinction
towards UV wavelengths. These observations suggest that the UV properties of the PAH are quite
different from those adopted here.
 
\section{THE FAR-IR SPECTRUM AND THE PRESENCE OF "COLD DUST"}

Dust model parameters were solely derived from
fitting the DIRBE $3.5-240\ \mu$m data, and Figure 3 shows the model
spectrum extended to FIRAS wavelengths. The figure shows that even without any attempts
to fit the FIRAS data, the model produces a good agreement with the $240-500\ \mu$m
spectrum. Figure 5 shows the fractional residual emission, defined as
$[F(FIRAS)-F(model)]/F(model)$, as a function of wavelength. The residual exhibits a
systematic trend in the $100-500\ \mu$m wavelength region,
consistent with an underestimate in the calculated dust temperatures. To demonstrate the
sensitivity of the far-IR spectrum to changes in dust temperatures, we plotted in the
figure the fractional residual expected if we had attempted to fit a spectrum with an
intrinsic temperature of 17.5 K with a model at 16 K. The figure shows that such a
difference could account for the residual emission in the $240-500\ \mu$m wavelength regime.
Such a temperature increase would, for example, require the intensity of the LISRF to be
increased by a factor of 1.5, or alternatively, require the ratio of the UV-optical to IR
emissivity of the dust to be increased by the same factor.

Persistent in the observations is, however, an excess of emission at wavelengths $\gtrsim
500\ \mu$m above the calculated model spectrum. In each of the FIRAS channels
the excess is only a 1 $\sigma$ detection, but considering the similarity of the trend in
all the $500-1000\ \mu$m channels, the excess gains significance. It was previously
detected in the FIRAS data by Wright et al. (1991), Reach et
al. (1996),  and may also be present in the residuals between the fit of a dust
spectrum, characterized by a single-temperature of T=17.5 K and a $\nu^2$ emissivity law, to
an HI-correlated DIRBE and FIRAS Galactic emission spectrum (see Boulanger et al. 1996,
Figure 3). The nature of this excess emission component is still uncertain. We emphasize
that this emission component should not be confused with the much stronger one found at high
latitude by Reach et al. (1995), which is claimed by Puget et al. (1996) to be of extragalactic
origin.  

Assuming that the excess is not an instrumental effect, it should be of a Galactic origin,
since it was derived by correlating the FIRAS data with a template of Galactic 100 $\mu$m
emission. As a result, any extragalactic component such as the one reported by
Puget et al. (1996) would not be present in this data. Several explanations have been
suggested for a Galactic origin  of this component (\cite{w91},
\cite{w93}, \cite{r95}): (1) the excess could be due to a far-IR feature
in the dust emissivity; or (2) it could be due to a Galactic population of cold ($T\approx
6\ K$) dust particles. Their low temperature is not likely the result of shielding from
the interstellar radiation field, requiring enhanced dust cooling
mechanisms such as expected for fractal dust grain, or needles. Recent laboratory results on
the millimeter absorptivity of silicate grain materials (\cite{agl96}) show that the opacities
of these grains is significantly larger than obtained by extrapolating the Draine-Lee values
with a
$\lambda^{-2}$ emissivity law.

\section{SUMMARY}

We presented the {\it COBE} $3.5-1000\ \mu$m spectrum of the diffuse ISM at
latitudes $|b|\ge 45^o$ (Fig.~1) and showed that the near-IR intensity ratios cannot be
produced by any grain size distribution of bare silicate or graphite grains (Fig. 2). In
particular, we find that the 3.5-to-4.9 $\mu$m intensity ratio is consistent with PAHs
being the dominant emission component at these wavelengths. We used the DIRBE data to
derive an interstellar dust model that consists of PAHs and bare silicate and graphite
grains, stochastically heated by the local interstellar radiation field (LISRF). In
addition, we used the [C II] 158 $\mu$m line observed by the FIRAS to derive the 
$C^+/H$ abundance in the diffuse high latitude ISM. The results of this paper can be
briefly summarized as follows:
\newline
\newline
1) The dust model provides a very good fit to the $3.5-1000\ \mu$m diffuse ISM spectrum (see
Figure 3). Dust parameters are summarized in Table~3. The model requires essentially all the
available Mg, Si, and Fe, and $\sim$ 15 \% of the available O to be locked up in silicate
grains. About 20\% of the total carbon is locked up in PAHs (C/H = $(7.3\pm 2.2)\ 10^{-5}$) and
about 60-70\% is locked up in graphite (C/H $\approx (2.5\pm 0.8)\ 10^{-4}$). The abundance of
PAHs in our model is higher by a factor of $\sim$2 compared to that derived in previous models
of low latitude cirrus.
\newline
\newline
2) The energy radiated by the
PAHs per unit H atom is $1.6\ 10^{-31}\ W/H\ atom$, and the energy absorbed from the
LISRF per unit carbon atoms in PAHs is $2.6\ 10^{-27}\ W/C\ atom$. The ratio between
the two quantities provides an independent estimate of the C/H ratio locked up in PAHs.
The resulting value, $(C/H)_{PAH}=6.1\ 10^{-5}$, in very good agreement with the PAH abundance
required to account for the IR emission.
\newline
\newline
3) The detection of the [C II] 158 $\mu$m line by the FIRAS provides important
constraints on the gas phase abundance in the cold neutral medium (CNM). We estimate
the $C^+/H$ ratio to be between $\sim(0.5-1)\ 10^{-4}$ (see Appenix A.3), constituting
$\sim$ 10-20\% of the cosmic abundance of carbon.
\newline
\newline
4) The total amount of carbon in PAHs, graphite grains, and in the gas is $C/H \sim
(4.0\pm1.2)\ 10^{-4}$. This value is consistent with the {\it cosmic} carbon abundance, but
barely consistent with the {\it interstellar medium} value of $C/H=(2.25\pm 0.5)\ 10^{-4}$.
\newline
\newline
5) The model provides a good fit to the FIRAS spectrum in the 240 to 1000 $\mu$m
wavelength regime. Detailed comparison of the residual spectrum (Figure 5) shows a
systematic trend consistent with an underestimate of the interstellar dust temperature.
The figure also shows the presence of an excess of emission at $\lambda\gtrsim 500\ \mu$m above
that predicted by the model. This excess emission correlates with the template of
Galactic 100 $\mu$m emission, and is therefore of Galactic origin. The nature of this
excess emission is still unresolved. 
\newline
\newline
6) Our model provides generally a good fit to the average interstellar extinction curve
(see Figure 4a). However, it produces an excess of extinction at FUV wavelengths
(Fig. 4b). This excess is mainly due to the higher abundance of PAHs in the model, and the fact
that we used Draine-Lee graphite particles with rising extinction in the FUV in combination
with PAHs which have similar optical properties at FUV wavelengths. A fraction of the excess
extinction may be real, reflecting a trend of increasing UV extinction with
latitudes. The trend may be the result of an increasing PAH abundance at high Galactic
latitudes. 

\acknowledgments
J. Ballester acknowledges the support of the NASA JOVE Program. E. D. acknowledges the 
hospitality
of the Institut d'Astrophysique Spatiale in Orsay where the final revision to the manuscript
was made. We also acknowledge helpful conversations with Xavier D\'esert,  Francois Boulanger,
Bruce Draine, Derck Massa, John Mathis, and Jean-Loup Puget.

\appendix
\section{CHARACTERIZATION OF THE DUST AND PAH MOLECULES}
\subsection{Heat Capacities}

The heat capacity plays a special role in determining the spectrum of stochastically
heated particles. Small values of $C(T)$ will accelerate the cooling around $T$, reducing
the flux emitted at the appropriate wavelengths. In contrast, large heat capacities,
especially at very low temperatures, may result in an excess of long-wavelength emission.

At sufficiently high temperatures, the heat capacity of the various PAHs can be
calculated using  the group additive method  described by Stein, Golden, and Benson
(1977; hereafter SGB). In this method, a PAH is  decomposed into four carbon groups
characterized by the composition of their nearest neighbors. For simplicity, we designate
these atomic groups as groups $A-D$. Group $A$ consists of the
$CÐH$ group in benzene $(C_6H_6)$; group $B$ of unsubstituted carbon in naphthalene
$(C_{10}H_8)$; group $C$ is found in phenanthrene
$(C_{14}H_{10})$, and group $D$ consists of interior carbon, identical to that
encountered in a monolayer of graphite . For example, chrysene ($C_{18}H_{12}$ consists
of 12 members from group $A$, 2 from group $B$, and 4 from group $C$ (it has no interior
carbon atoms; see, for example, Fig. 22 of Allamandola, Tielens, \& Barker (1989). The
number of members from each group in a given PAH are not independent, and one can derive
several simple relationships between them and $N_c \equiv N_A+N_B+N_C+N_D$, the total
number of carbons in a PAH. For a fully hydrogenated PAH, the number of hydrogen atoms is
given by
$N_H =
\sqrt{6N_c}$  (see also \cite{o86}), and $N_A = N_H$ from the group definition.
Empirically, we find that $N_B + N_C = N_A - 6$. To these four groups we added a group
$A^*$ to represent a dehydrogenated $A$ group, so that $N_{A^*} = (1 - f_H) N_A$, where
$f_H$ is the number of filled hydrogen sites ($f_H = 0$ for a completely dehydrogenated
PAH).

The heat capacity of a given PAH at temperature $T$, $C_{PAH}(T)$, is then given in the
group additivity technique by the sum:
\begin{equation}
\begin{array}{ll} 
C_{PAH}(T)= & f_HN_AC_A(T) + (1 - f_H)N_AC_{A^*}(T)+ N_B C_B(T) \\ 
            & + N_C C_C(T) +N_D C_D(T)
\end{array}
\end{equation}
where the N's are the number of group members in the PAH. Following SGB, we
assumed that $C_C$ is equal to $C_B$, and substitution of the values of the $N$'s in the
equation for $C_{PAH}(T)$, gives for a completely hydrogenated PAH ($f_H=1$)
\begin{equation}
C_{PAH}(T) = N_c \Big\{\ {\sqrt{6\over{N_c}}\ [C_A+C_B-2C_D] + {6\over{N_c}}
[C_D- C_B] + C_D\ }\Big\}
\end{equation}

The equation shows that for sufficiently large values of $N_c$, the heat capacity becomes
equal to the graphitic limit. The heat capacity of the
various groups in the 300 to 3000 K temperature range is given by Stein (1978) and Stein
\& Fahr (1985). For $C_{A^*}$ we used the heat capacity of phenyl radical [designated as
$C_B-(\cdot)$ by Stein and Fahr (1985)].

The resulting values for $C_A, C_B,$ and $C_D$ can be written as polynomials in
temperature: 
\begin{equation}
C(J\ mole^{-1}\ K^{-1})\ =\ \sum_{n=0}^{6} a_nT^n, 
\end{equation}
where for $C_A$:
\newline
$\{a_{0-6}\} = \{-1.23,\ 4.9\ 10^{-2},
\ 2.07\ 10^{-5},\ -6.93\ 10^{-8},\ 4.85\ 10^{-11},\ -1.44\ 10^{-14},\ 1.57\ 10^{-18}\}$
\newline
 for $C_B$:
\newline
 $\{a_{0-6}\} = \{-7.02,\ 48.87\ 10^{-2},
\ -1.11\ 10^{-4},\ 7.98\ 10^{-8},\ -3.33\ 10^{-11},\ 7.42\ 10^{-15},\ -6.79\ 10^{-19}\}$
and for $C_{A^*}$:
\newline
 $\{a_{0-6}\} = \{4.01,\ 1.62\ 10^{-2},
\ 4.89\ 10^{-5},\ -1.20\ 10^{-7},\ 1.07\ 10^{-10},\ -4.41\ 10^{-14},\ 6.91\ 10^{-18}\}$

Since the group additive method breaks down at PAH temperatures below
$\sim 300\ K$, the heat capacity at these temperatures was fitted with that of
graphite, scaled to smoothly merge with that calculated by the group additive method.
The resulting value for the heat capacity in $erg\ K^{-1}$, valid for T $\leq$ 2000 K, is given
by:
\begin{equation}
log_{10}[C_{PAH}(T)/N_c]=-21.26 + 3.1688\ log_{10}T\ - 0.401894\ (log_{10}T)^2
\end{equation} 
In the limit of small carbon atoms, the PAH heat capacity reproduces tabulated values of
naphtalene ($C_{10}H^8$), for which $C(T)=8C_A(T)+2C_B(T)$, and anthracene
($C_{14}H_{10}$), for which $C(T)=8C_A(T)+2C_B(T)$. For large enough PAHs, the group
additive method matches the heat capacity of graphite. The value of $C_{PAH}(T)$ in eq. (A4)
is essentially identical to the one presented by Puget
\& L\'eger (1989).

For graphite, the heat capacity in the $10 - 2000\ K$ temperature is given in an
easily integrable form by (\cite{mark73}):
\begin{equation}
C_{grf} (T) = 2.2\times 10^7\sum_{n=3}^{5} a_nT^n/ \sum_{n=0}^{5} b_nT^n
\ \ \ [\ erg\ K^{-1}cm^{-3}]
\end{equation}
where the coefficients are given by: $\{a_3, a_4, a_5\} = \{0.10273,\
4.4354\ 10^{-2},\ 2.2124\ 10^{-4}\}$, and
$\{b_0, b_1, b_2, b_3, b_4, b_5\}=\{1.0003\ 10^{-12},\ 3.6909\ 10^4,\ 1129.71,\ 30.4420,\
1.2888\ 10^{-2},\ 1.0000\ 10^{-4}\}$.

Silicate heat capacities were adopted from the values given by Draine \& Anderson
(1985). Silicate and graphite heat capacities were corrected by a factor of $(1-2/N)$,
where $N$ is the number of atoms in the grain, to take into account that the heat
from photon absorption events gets only distributed in the vibrational modes of the
solid (\cite{gd89}). No such correction was needed for the PAHs since in the small particle
limit their heat capacities reproduce that of observed molecules.   

\subsection{Cross Sections}

The cross section for graphite and silicate dust particles is given by:
\begin{equation}
\sigma(a,\lambda) = \pi a^2 Q(a,\lambda)
\end{equation}
where $Q(a,\lambda)$ is the dust absorption efficiency. Values of
$Q(a,\lambda)$ were calculated from Mie theory using the dust optical constants from
Draine \& Lee (1984).

The PAH cross sections used in this paper were adopted from D\'esert, Boulanger, \& Puget
(1990). In their model, the PAH cross section $\sigma_{PAH} (\lambda)$ is
represented by a sum of UV-visual, IR continuum, and IR line terms, as follows:
\begin{equation}
\sigma_{PAH}(\lambda)=\sigma_{UV-vis}(\lambda)+\sigma_{IRc}(\lambda)+\sigma_{IRl}(\lambda)
\end{equation}
where 
\begin{equation}
\begin{array}{lll} 
\sigma_{UV-vis}(\lambda) & = &10^{-18} N_c\ [p_1f_v(x)+p_2f_u(x)]\ {\cal C}({x\over
x_c})\ \ \       cm^2\\
\sigma_{IRc}(\lambda) & = &{3.3\times 10^{-20}\over \lambda(\mu m)} N_c\ 
exp(-{\lambda_m\over\lambda})\ \ \  cm^2\\
\sigma_{IRl}(\lambda) & = &\sum_j{\sigma_j\times exp(-{(\lambda-\lambda
_j)^2\over{2(\Delta\lambda_j)^2}})}
\end{array}
\end{equation} 
$N_c$ is the number of C-atoms in the PAH, $x\equiv 1/\lambda(\mu m)$, $p_1=4.0\
\  $and$ \  p_2=1.1$ are numerical parameters,
$x_c\equiv {12.5\over {a_{PAH}(\hbox{\AA})}},\ a_{PAH}(\hbox{\AA})=10 \sqrt {N_c/120}$ is the
radius of a PAH, and $\lambda_m =\ 10\ \mu$m. 
\newline
The functions $f_u, f_v,$ and $C(y)$ are given by:
\begin{equation}
\begin{array}{llll} f_u(x)&=&(x-5.9)^2\ (0.1x+0.41) & $for$\ x\ge 5.9\ \mu m^{-1}\\ &=&0 &
$otherwise$\\ f_v(x)&=&1.0 & $for$\ x\ge x_l\equiv 4\ \mu m^{-1}\\ &=&x^2\ (3x_l-2x)/x_l^3
&$for$\ x\le x_l
\end{array}
\end{equation}

\begin{equation} 
C(y\equiv x/x_c)=\pi^{-1}\ \arctan (10^3\ (y-1)^3/y)+0.5
\end{equation}

The last expression in eq. (A8) assumes that the PAH spectral features are Gaussian in
shape. The cross section is represented as a sum over the PAH feature $j$, where
$\sigma_j$ is the value of the cross section at the line center
$\lambda_j$, \{$\sigma_j/10^{-21}\ cm^2$\}=\{$35N_H,\ 4.1N_c,\ 2.9N_c,\ 3.0N_H,
\ 47N_H$\}, for, respectively the 3.3, 6.2, 7.7, 8.6, and 11.3 $\mu$m bands (L\'
eger, d'Hendecourt, \& D\'efourneau 1989; hereafter LHD). The
$\Delta \lambda_j$'s are the effective width of the $j$ features, satisfying the condition
that the integral of the cross section over the line equals  $ {\cal A}_j\equiv
\lambda_j(\mu m)^2\times f_j/1.13\times 10^{20}$, where
$f_j$ is the dimensionless oscillator strength (LHD). The linewidths in our
representations are then given by: 
$\Delta \lambda_j(\mu m) = 3.53\times 10^{-17}\ f_j\lambda_j(\mu m)^2/\sigma_j(cm^2)$.
The resulting values for  $\Delta \lambda_j(\mu$m) are \{0.0165, 0.070, 0.274, 0.157,
0.115\} at \{$\lambda_j(\mu$m)\}= \{3.3, 6.2, 7.7, 8.6, 11.3\}, respectively. Note that
our values for
$\Delta \lambda_j(\mu m)$ are smaller by a factor of
$\sqrt{2\pi}$ from those quoted by LHD because our choice of the functional
representation of the features. The actual functional representation is irrelevant for
broad band photometry (such as the DIRBE data) as long as the condition that 
${\cal A}_j\equiv \int{\sigma_j(\lambda) d\lambda}$ across a feature is satisfied.


\subsection{The Carbon Gas Phase Abundance in the Neutral Medium}
 The average high-latitude ISM
spectrum calculated in the model represents the IR emission integrated over
$\it{all}$ gas phases along the line of sight (see eq. 1), including in addition to the
H I, any high latitude molecular or ionized gas that correlates with the 100 $\mu$m
emission.
However, we do not expect emission from molecular gas to be a significant contributer to
the IR spectrum from the average ISM gas. The high-latitude CO
survey of Magnani, Blitz, \& Mundy (1985) found that the surface filling factor of the
high-latitude molecular gas is low, with estimates ranging from $\sim$ 6 to 10\% (see
review by \cite{m94}). Rarely, molecular gas may be the dominant phase in select cirrus
clouds, with MBM 53-55 an extreme example (see \cite{dar96}).

On the other hand, the warm ionized medium (WIM; a low density ionized gas
that gives rise to pulsar dispersion measures, and diffuse H$\alpha$ emission; e.g.
Reynolds 1990), can be a significant contributor to the
observed IR emission. Pulsar dispersion measures show that at high-latitudes, the
$H^+/H\ I$ column densities are $\sim\ 0.2 - 0.6$ (e.g. Reynolds 1991). Assuming
that the dust in the WIM is exposed to a radiation field similar to the LISRF and have a
similar dust-to-gas mass ratio as in the general ISM, we get that at most one third of the
observed IR intensity can originate from the ionized gas.

We therefore assume that most of the IR emission originates from the neutral
medium, and that almost all of the gaseous carbon in the average high-latitude ISM is in
the form of $C^+$, which dominates the contribution of heavy ions to the electron density.
With an ionization potential of 11.26 eV it is easily ionized by the diffuse interstellar
radiation field, which is not hard enough to eject a second electron.
$C^+$ has a ground-state fine-structure line at
$158\ \mu$m, which is collisionally excited by electrons, H I and H$_2$. The amount of
gaseous C can therefore be estimated from $I(158\ \mu$m) observations. The 158 $\mu$m
intensity can be calculated from the FIRAS observations
(see Figure 1). Subtracting the dust continuum intensity calculated by the model we derive
a value of
$4\pi I(158\ \mu m)/H = 1.45\times 10^{-33}\ W/H\ atom$, a value lower by about a
factor of $\sim 2$, compared to the $|b|\ge15^o$ line intensity given by Bennett et al.
(1994).  

Most of the contribution to the observed $I(158\ \mu$m) intensity comes from the neutral
gas (\cite{ben94}), instead of the WIM. The argument presented by Bennett et al. was
simple: any
[C II] 158 $\mu$m emission from from the WIM will be accompanied by
H$\alpha$ emission. The relation between the two quantities is given by (Reynolds
1992):
\begin{equation} I(158\ \mu m)\simeq 1.45\times T_4^{0.57} ({C^+/H^+\over 3.3\ 10^{-3}})\
I(H\alpha)
\end{equation} where
\begin{equation} I(H_{\alpha})\simeq 8.7\times 10^{-11} T_4^{-0.92}\ EM\ \ \ W\ m^{-2}\
sr^{-1}
\end{equation} 
and where $T_4=T/10^4\ K$, and $EM\equiv \int n_e^2 ds$ is the emission
measure in units of $cm^{-6}\ pc$. Values of EM along several high latitude lines of
sight are typically $\sim 1-2\ cm^{-6}\ pc$, giving upper limits for the $I(158\
\mu$m) intensity of 
$\sim (2-3)\times 10^{-10}\ W\ m^{-2}\ sr^{-1}$, assuming that all the carbon
is in the gas phase, and singly ionized. This intensity is significantly below the FIRAS
observed value of $\sim 2\times 10^{-9}\ W\ m^{-2}\ sr^{-1}$
(\cite{w91}, \cite{ben94}). A similar conclusion was reached by Bock et al. (1993), who
derived a value of $I(158\ \mu m)\approx (9\pm3.5)\times 10^{-11} W\ m^{-2}\
sr^{-1}$ for the line intensity from the WIM by extrapolating the  $I(158\ \mu$m)
versus $N_{H\ I}$ correlation to $N_{H\ I}=0$.
 
Since most of the  $I(158\ \mu$m) emission originates from the neutral medium, we can
can use the observed line intensity to derive an estimate of
the $[C^+]/[H\ I]$ ratio in this  medium, if the physical conditions in the medium
(temperature, density, ionization state) are known. Designating the  upper and lower energy
levels of the
$158\
\mu$m transition as levels 2 and 1, respectively, the line intensity (in erg
$s^{-1} cm^{-2} sr^{-1}$) is simply the integral of the line cooling rate along the line
of sight (e.g. \cite{O89}), and can be written as:
\begin{equation} I(\nu)={A_{21}h\nu
\over{4\pi}}N_{C^+}({g_2\over g_1}) e^{-h\nu/kT} { \sum_j({n_j\over
n_{c,j}})\over{1+\sum_j({n_j\over n_{c,j}})}}
\end{equation}
where $h\nu = 1.26\ 10^{-14} erg$ is the energy of the transition, $N_{C^+}$ is the
column density of
$C^+$ atoms along the line of sight,
$A_{21}=2.36\ 10^{-6}\ s^{-1}$ is the Einstein A-coefficient for the 2 $\rightarrow$ 1
transition, $g_2=4$ and
$g_1=2$ are, respectively the statistical weights of levels 2 and 1, $n_j$ is the number
density of the colliding species, and $n_{c,j}\equiv A_{21}/q_{21}$ is the critical
density, where
$q_{21}=8.629\ 10^-6\ \Omega_j/g_2T^{1/2} cm^3\ s^{-1}$ is the collisional de-excitation
rate, and
$\Omega_j$ is the dimensionless collision strength with species "j". For electronic
collisions, $\Omega
\approx 2.8$ in the $\sim$ 5000 to 20,000 K temperature interval, and $\Omega \approx
1.8$ for temperatures below
$\sim 1000\ K$ (Hayes \& Nussbaumer 1984). For collisions with H, $\Omega = 0.00292$
(Launay \& Roueff 1977). Normalizing the observed line intensity to the observed
H I-column density along the line of sight, eq (13) can be used to calculate the
$C^+$-to-H I abundance ratio, giving:
\begin{equation} {[C^+]\over [H\ I]}    ={ (g_1/g_2)\ e^{h\nu/kT}\over {A_{21}h\nu}}\
[{4\pi I(\nu)\over N_{H\ I}}] / \sum_j({n_j\over n_{c,j}})      
\end{equation} where we have made the approximation that $n_j\ll n_{c,j}$.

Most of the line emission originates from the cold neutral medium (CNM) which has
typical parameters \{$n_H, T$\} =
\{$20-80\ cm^{-3}, 40-100\ K$\}, with pressures $P/k\approx 2500\ cm^{-3}\ K$ (e.g. Heiles
1994). This medium is kept partially ionized by high-energy
cosmic rays (CR). The electron density,
$n_e$ maintained by a CR ionization rate
$\zeta_{CR}$ (in $s^{-1}$) is given by (\cite{s78}):
\begin{equation} 
n_e={n_i\over 2}\ [1+(1+{4\zeta_{CR}\ n_H\over \alpha^{(2)} n_i^2} )^{1/2}
]
\end{equation} 
where $\alpha^{(2)}\approx 6.2\ 10^{-11}\ T^{-1/2}\ cm^3\ s^{-1}$ is the
partial H$-$recombination coefficient to the n$>$1 levels of hydrogen, and $n_i$ is the total
number density of photo-ionized heavy elements.

Equations (A14) and (A15) can be solved for ${[C^+]/[H\ I]}$ as a function of the
parameters of the CNM, the CR ionization rate $\zeta_{CR}$, and the observed $I(158 \mu
m)/N_{H\ I}$ ratio. Estimates of $\zeta_{CR}$ range from $\approx 10^{-16} - 10^{-18}\ 
s^{-1}$. For the
$C^+$ cooling rate per H-atom of $4\pi I(158 \mu m)/N_{H I} = 1.45\ 10^{-33}\ W/H\ atom$
found earlier, and an adopted pressure of $P \approx 2000 - 3000\ cm^{-3} K$, we get  a
$[C^+]/[H I]$ ratio of:
\begin{equation}
{[C^+]\over [H I]}\approx (0.5 - 1)\times 10^{-4}
\end{equation}
For a mean atomic weight of $\mu = 1.42$, the resulting $C^+$-to-gas mass
ratio,
$Z_{C^+}\approx (4 - 9)\times 10^{-4}$. 


\begin{deluxetable}{llll}
\tablecaption{Dust Spectra/color temperatures in the Average ISM and Selected Cirrus
Clouds\tablenotemark{a}}
\tablehead{
 & \multicolumn{3}{c}{$I(\lambda)/I(100\ \mu$m)\ /\ $T_{color}$ (K)} \nl
\colhead{$\lambda(\mu$m)} & \colhead{average ISM\tablenotemark{b}} 
  & \colhead{Cloud 1\tablenotemark{c}} & \colhead{Cloud 2\tablenotemark{d}} 
  }
\startdata 
3.3	 & $0.00168$               & 0.00161        & 0.00124  \nl  
4.9  & $0.00252\  /550$        & 0.00074 /1500  & 0.00186  /500  \nl 
12	  & $0.0464\ \ /230    $    & 0.0343\ \ /\ \ 200    & 0.0600 \  /210  \nl 
25   & $0.0389\ \ /210      $  & 0.0283 \ /\ \ 210  & 0.0633 \  /190  \nl 
60   & $0.160\ \  \  /\ \ 70  $ & 0.114\ \ \ /\ \ \ \ 70 & 0.178 \ \   /\ \ 75   \nl
100  & $1.0000\ \ /\ 23     $  & 1.0000  \ /\ \ \ 44   & 1.000\ \ \  /\ 24   \nl 
140  & $1.93\ \ \  \ /\ \ 18  $  & 2.53\ \ \ \ /\ \ \ 16    & 2.45\ \ \ \  /\ \ 16   \nl 
240  & $1.28\  \ \ \ /\ \ 19  $  & 2.51\ \ \ \ /\ \ \ 16   & 1.36\ \ \  /\ \ 21   \nl
\enddata
\tablenotetext{a}{Fluxes are color-corrected and given in units of 
$0.7\ MJy\ sr^{-1}/N_{H\ I}(10^{20}\ cm^2)$. Numbers in parethesis  in each row "j" are
the color temperatures determined from the $I(\lambda_{j-1})/I(\lambda_j)$ flux ratios
for a $\lambda^{-2}$ dust emissvity law}
\tablenotetext{b}{Fluxes represent Galactic average IR fluxes integrated over all gas
phase  components that spatially correlate with the observed H I emission (see \S 2 for
more detail)}
\tablenotetext{c}{average spectrum of the cloud complex MBM 53, 54, and 55 (see \S 2 for
more details)}
\tablenotetext{d}{average spectrum of cloud A (Low et al. 1984)}  
\end{deluxetable}

\clearpage
\begin{deluxetable}{lll}
\tablecaption{Dust Model Parameters}
\tablehead{
\colhead{Parameter} &\multicolumn{2}{c}{Value} \nl
\colhead{ } & \colhead{Model A} & \colhead{Model B} \nl 
  }
\startdata
\underline{PAH parameters} & & \nl
$N_{c1}$	   & 20               &  20            \nl
$N_{c2}$    & variable         & variable       \nl
$\gamma_p$  & variable         & variable       \nl
$Z_{PAH}$   & variable         & variable        \nl
  & & \nl
\underline{graphite parameters} & \underline{model A}& \underline{model
B}\tablenotemark{a}\nl
$a_{min}(\mu$m) & variable & $1.29\times 10^{-4}N_{c2}^{1/3}$   \nl
$\gamma_1$ 					& -3.5 				& $3\gamma_p+2$  \nl
$a_b(\mu$m)	    & 0.1      & variable       \nl
$\gamma_2$      & -3.5     & -3.5           \nl
$a_{max}(\mu$m) & 0.25     & 0.25           \nl
$Z_{grf}$       & variable & variable       \nl
 & & \nl
\underline{silicate parameters} & & \nl
$a_{min}(\mu$m) & 0.0050   & 0.0050      \nl
$\gamma$        & -3.5     & -3.5        \nl
$a_{max}(\mu$m) & 0.25     & 0.25        \nl
$Z_{sil}$       & variable & variable    \nl
\enddata
\tablenotetext{a}{In model B, $a_{min}$ is calculated so that the number of carbon atoms
in a spherical graphite grain of that radius is equal to $N_{c2}$, the number of carbon
atoms in the largest PAH, 	and $\gamma_1$ is calculated so that $dn(N_c)/dN_c$ is
continuous across $a_{min}$, the PAH-graphite radius boundary}  
\end{deluxetable}

\clearpage
\begin{deluxetable}{llllll}
\tablecaption{Summary of Dust Model Parameters\tablenotemark{a}}
\tablehead{
\colhead{Model Parameters\tablenotemark{b}} & \multicolumn{5} {c}{Average Diffuse ISM
Dust}  \nl
\colhead{ } & \multicolumn{2} {c} {Model A} & \colhead{ } &\multicolumn{2} {c} {Model B}
\nl 
 }
\startdata
\underline{\it{PAH}}    &  & & & &  \nl
$N_{c1}$                                 & 20        & 20     &  & 20       &  20    \nl
$N_{c2}$	                                & 100       & 171    &  & 100      & 163    \nl
$\gamma_p$                               & -1.67     & -2.33  &  &- 1.78    & -2.67   \nl
$Z_{PAH}$ 																															& 0.00062   & 0.00061 &  & 0.00060 & 0.00054  \nl
  & & & & & \nl
\underline{\it{Graphite}}	 &  & & & &  \nl
$a_{min}(\mu$m)   				                   & 0.0005   &  0.0005 &  & 0.0006   & 0.0007 \nl
$\gamma_1$                               & -3.5     &  -3.5   &  & ---      & -6.00  \nl
$a_b(\mu m)$                             & 0.10     &  0.10   &  & 0.0006   & 0.0012 \nl
$\gamma_2$                               & -3.5     &  -3.5   &  & -3.5     & -3.5    \nl
$a_{max}(\mu m)$    																					& 0.25     &  0.25   &  & 0.25     & 0.25   \nl
$Z_{grf}$ 																															& 0.0021   & 0.0020  &  & 0.0022   & 0.0023 \nl
  & & & & & \nl
\underline{\it{Silicate}}	  & & & & & \nl
$a_{min}(\mu m)$ 																								& 0.0050   & 0.0050  &  & 0.0050   & 0.0050 \nl
$\gamma$          																							& -3.5     & -3.5    &  & -3.5     & -3.5   \nl
$a_{max}(\mu m)$  																							& 0.25     & 0.25    &  & 0.25     & 0.25   \nl
$Z_{sil}$																															 & 0.0053  & 0.0058    &  & 0.0046  & 0.0042 \nl
  & & & & &  \nl
\underline{\it{Solid Phase Abundances}} & & & & & \nl
$Z_{carbon}$	                		          & 0.0027  & 0.0026 &  & 0.0028  & 0.0028 \nl
$Z_{dust}$ 					                         & 0.0080  & 0.0083 &  & 0.0074  & 0.0070 \nl
 & & & & & \nl
\underline{\it{Gas Phase Carbon Abundances\tablenotemark{c}}} & & & & & \nl
$Z_{C^+}$              & \multicolumn {2} {c} {0.0004-0.0009}
& & \multicolumn {2} {c} {0.0004-0.0009}      \nl
\enddata
\tablenotetext{a}{See Table 2 for the definitions of Models A and B}
\tablenotetext{b}{Dust-to-gas mass ratios are given in units of  ${N_H\over
N_{H I}}$. For comparison, the cosmic abundances of refractory elements that can be locked
up in dust are:
$Z_{carbon}(\odot)=0.0030;\   Z_{sil}(\odot)=0.0043$ assuming a silicate composition
of $\{MgSiFe\}O_4$. Elemental abundances were taken from Anders \& Grevesse (1989).}
\tablenotetext{c}{Calculated from the $I(158\ \mu$m$)/I(100\ \mu$m) flux ratio. See Appendix
 A.3 for more details}
\end{deluxetable}

\clearpage
\begin{figure}
\caption{The various dust spectra observed by DIRBE and FIRAS (the latter is given only
for the average ISM). The average spectrum of dust in the diffuse ISM normalized at $100\
\mu$m to a flux of
$0.7\ MJy\ sr^{-1}/N_{H\ I}(10^{20}\ cm^{-2})$. The other spectra have arbitrary
normalizations, and are offset from the diffuse ISM spectrum for sake of clarity.}
\end{figure}

\begin{figure}
\caption{A comparison of observed intensity ratios with calculated DIRBE intensity
ratios for PAHs and individual grains of various sizes and compositions. Intensity
ratios for the individual PAHs were calculated as a function of $N_c$, the number of
carbon atoms in the molecule, and for illustrative purposes converted to a radius using
the relation: $a$(\AA)=0.913$\protect\sqrt{N_c}$.}
\end{figure}
 
\begin{figure}
\caption{Fit of our interstellar dust model to the average dust spectrum in the diffuse
ISM for model A with $N_{c2}=100$. Details of the grain size distribution and relative
abundances of the various dust components can be found in the text (\S 4, and Table 3).}
\end{figure}

\begin{figure}
\caption{The average extinction for the diffuse interstellar medium derived from model A
with $N_{c2}=100$ (see Table 3), is compared with the observed average interstellar
extinction curve presented by Mathis (1990) renormalized to H I column densities.
Detailed of the comparison are presented in the text.}
\end{figure}

\begin{figure}
\caption{Residuals of between the FIRAS average ISM spectrum and model calculations. The
residuals show a trend in the $100 - 500\ \mu$m region consistent with an underestimate
in the temperature of the emitting dust. The smooth curve depicts the residuals expected 
with such an underestimate (see text for details). However, even with such a correction, the
figure shows a persistent excess of emission at $\lambda > 500\ \mu$m.}
\end{figure}

\end{document}